\documentclass[slac_one]{revtex4}
\usepackage{graphicx}
\usepackage{fancyhdr}
\begin{document}

\title{Neutrino Oscillation Measurements with IceCube} 


\author{C.~Rott (for the IceCube Collaboration$^*$)}
\affiliation{Center for Cosmology and Astro-Particle Physics, The Ohio State
University, 191 W. Woodruff Ave., Columbus, OH 43210, USA \\
$^*$http://icecube.wisc.edu}

\begin{abstract}

We present preliminary results for a neutrino oscillation analysis in progress 
on data collected with the IceCube 22-string detector during
2007 and 2008. 
The goal of this analysis is to measure muon neutrino disappearance as
a function of energy for a constant baseline length of the diameter of 
the Earth by studying vertically up-going muon neutrinos.
At this baseline disappearance effects
are expected to become sizable at neutrino energies below 100~GeV.
This energy range has not been previously explored with IceCube, however
due to IceCube's vertical geometry there is some sensitivity for 
this specific class of events.
Based on preliminary selection criteria, 
we show that IceCube has the potential to detect 
these events and we estimate the sensitivity to determining oscillation parameters.

\end{abstract}

\maketitle

\thispagestyle{fancy}


\section{IceCube Neutrino Observatory}

The IceCube Neutrino Telescope~\cite{IceCube} is a multipurpose discovery experiment 
under construction at the South Pole. It is currently about half
completed and is taking physics data. Upon completion in 2011, 
IceCube will instrument a volume of one cubic kilometer with 4800~Digital
Optical Modules (DOMs). These will be vertically spaced with 17~m separation at a depth between 1450~m
and 2450~m in the ice on 80~strings that are arranged in a hexagonal pattern
with an inter-string spacing of about 125~m. 
An additional six strings will be deployed in-between the 
hexagonal pattern at the center of IceCube in the 
deep very clean ice region with DOMs more densely spaced and equipped with high quantum
efficiency 
photomultiplier tubes. They will form together with the adjacent IceCube strings the 
Deep Core~\cite{Resconi:2008fe} subdetector. The first of these 
Deep Core strings will be deployed at the end of this year.

\section{Neutrino Oscillation Analysis}

Cosmic rays interacting with Earth's atmosphere generate a steady flux of 
secondary particles including muon neutrinos produced in kaon and pion decays.
These atmospheric neutrinos can be identified by IceCube through the
observation of Cherenkov light from muons produced in charged-current
interactions of the muon neutrinos with the Antarctic ice. 
The main difficulty in identifying these events stems from an 
overwhelmingly large down-going high energy muon flux produced in the
atmosphere, that penetrates the Earth several kilometers to IceCube depths and
can be misreconstructed as up-going events.

\begin{figure*}[h]
\begin{minipage}{18pc}
 \centering
 \includegraphics[width=16pc]{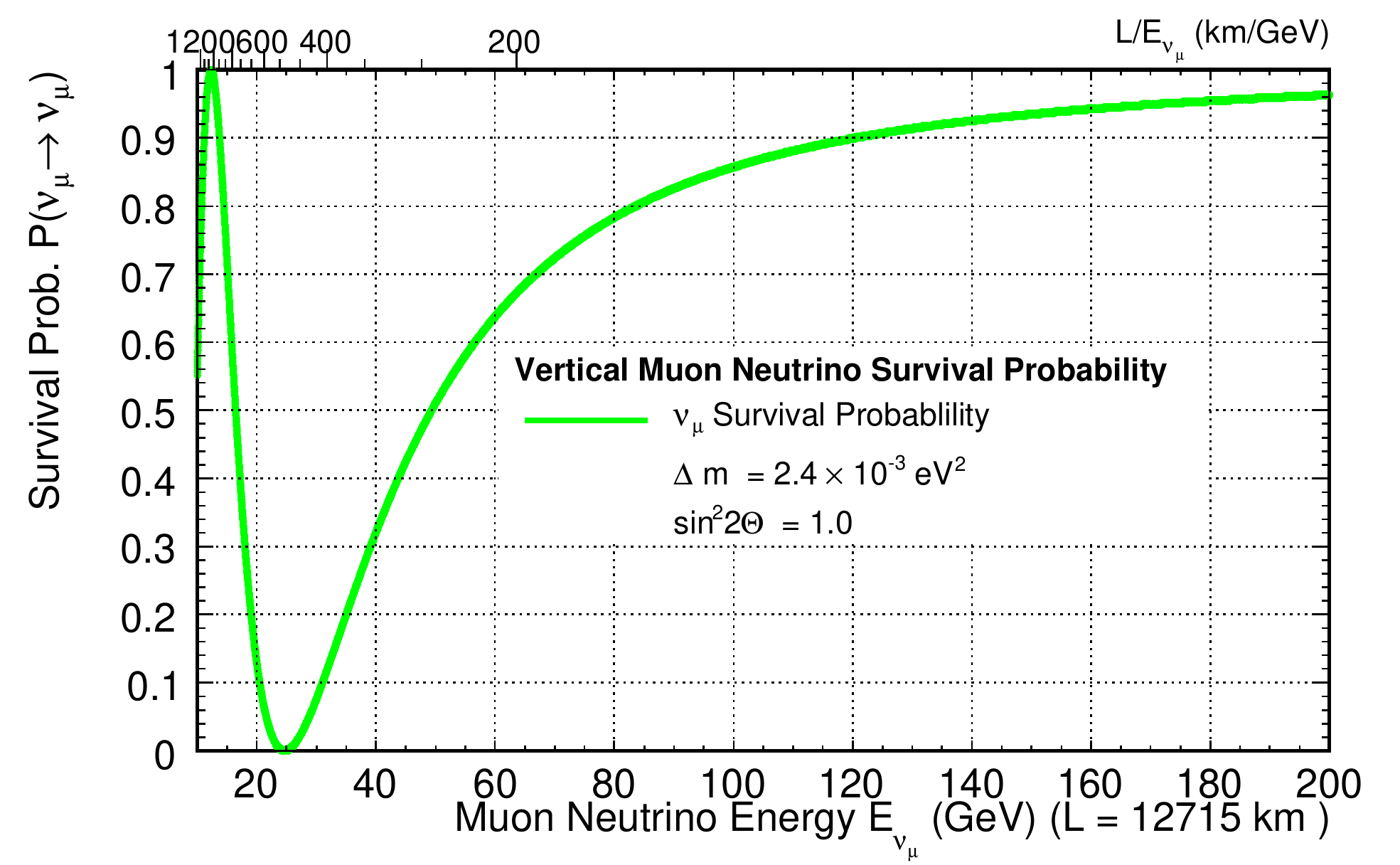}
 \caption{Muon neutrino survival probability under the assumption of effective
   2-flavor neutrino oscillations $\nu_\mu \leftrightarrow \nu_\tau$ as
   function of energy for vertically traversing neutrinos.} \label{plot:osc_LE}
\end{minipage}\hspace{2pc}%
\begin{minipage}{18pc}
 \centering
 \includegraphics[width=16pc]{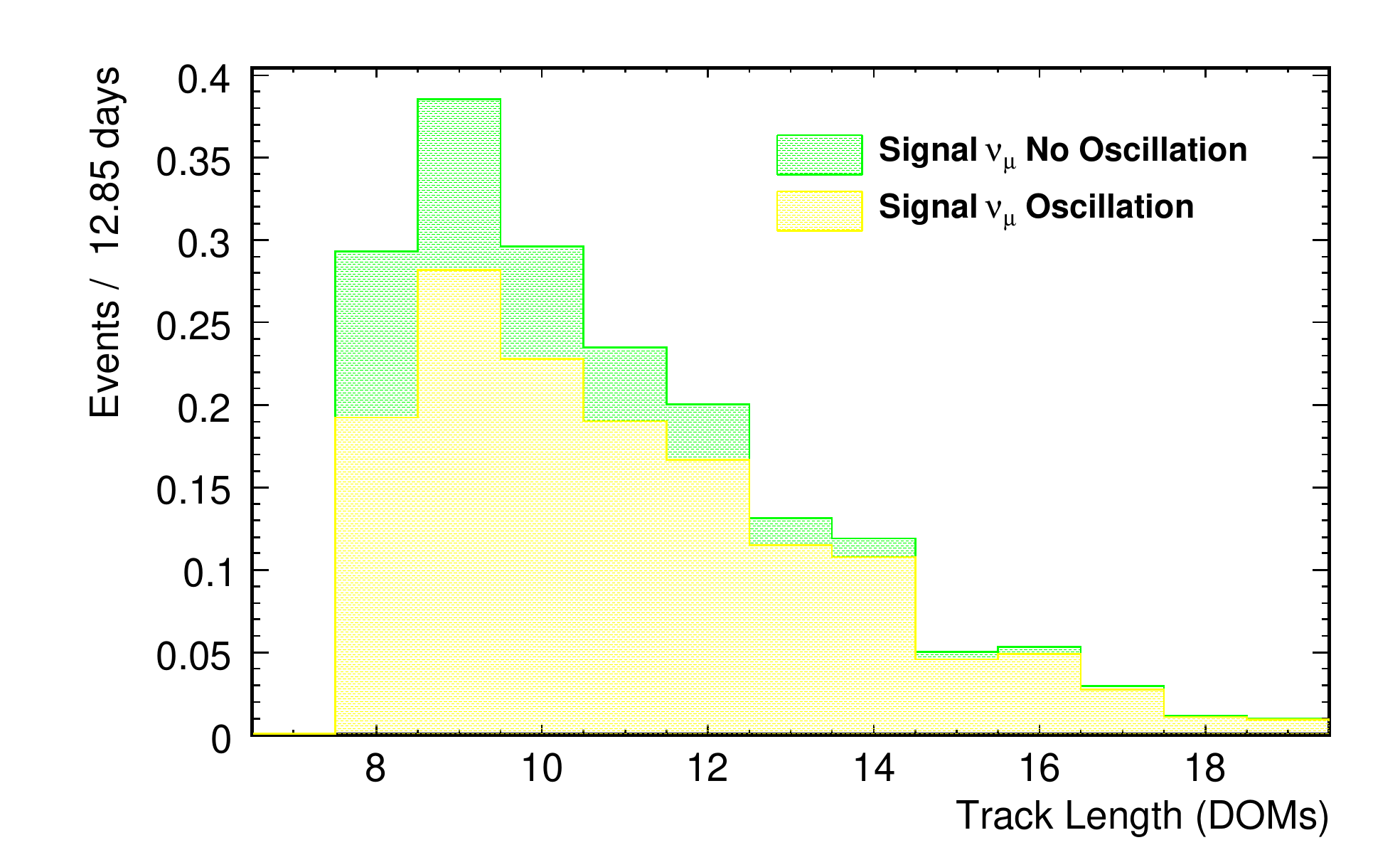}
 \caption{Signal distribution after selection criteria for the track length in number of DOMs hit
   along a string with and without oscillation effects taken into account.} \label{plot:trk_length}
\end{minipage}
\end{figure*}

In IceCube vertically up-going atmospheric neutrinos travel a distance of the 
diameter of the Earth (a baseline length L of about 12700~km). 
The survival probability for these muon neutrinos is shown in
Figure~\ref{plot:osc_LE} 
for maximal mixing and a $\Delta {\rm m}^2$ consistent with measurements 
by Super-Kamiokande~\cite{SK_DeltaM2} and
MINOS~\cite{MINOS_DeltaM2}. 
It illustrates the disappearance effect we intend to observe.
Oscillation effects become large for neutrino energies below 100~GeV.
This energy range is normally hard to access with IceCube; however, due to 
IceCube's vertical geometry, low noise rate, and low trigger threshold
(SMT8 - multiplicity eight DOM) the observation of neutrino oscillations through 
muon neutrino disappearance seem feasible. We require tracks to be 
vertical and near a string, so that the Cherenkov light can be sampled well
from even low-energy events.

Atmospheric neutrino oscillations have never been observed with AMANDA or
IceCube.
However, searches for non-standard oscillation mechanisms that lead to
observable differences at higher energies have been performed and 
constraints on these scenarios were placed~\cite{Ahrens:2007zzc}.

To search for muon neutrino disappearance we select vertical low energy muon neutrino
candidate events through the application of a series of consecutive selection criteria. First, events are
preselected from the data sample by a specific analysis
filter, 
which selects short track-like single string events. We then
require that the events are vertically up-going through a directionality selection,
which requires that the majority of time differences between adjacent DOMs are
consistent
with coming from unscattered Cherenkov radiation of a vertically up-going muon.
After these selection criteria the dataset is still
dominated by down-going muon background mimicking up-going events.
This background is estimated from CORSIKA~\cite{CORSIKA}
simulations and agrees well with data. 
Based on background and signal (atmospheric $\nu_\mu$ were generated with 
ANIS~\cite{neutrino_generator}) simulations we define
a tight set of selection criteria to further reduce the background. 
Thereafter, we reject all events in the available background MC sample, which has
the equivalent of about 1.45~days of livetime.
We expect 1.81 (1.42) signal events (with oscillation
effects taken into account) from atmospheric neutrinos in 12.85~days of
livetime,
which represents a small fraction of the IceCube 22-string dataset.
For these signal events the estimated expect distribution of the 
track length, which serves as an energy estimator and
works well at the energy range of interest as a muon travels roughly 5m/GeV, 
is shown in Figure~\ref{plot:trk_length}.
As expected, short tracks show larger disappearance effects.
The optimization and cross-check on the small subset of available data have been
performed in a blind manner. We observe three signal candidate events after
final selection, which is consistent with the predictions. 
This initial result confirms that we understand and model the
low-energy atmospheric neutrino region reasonably well. 
However, a larger MC background sample is needed estimate any possible
remaining background after tight selection criteria.
The analysis on the full dataset is in progress, including a larger background
MC sample and a detailed study 
of systematic uncertainties.

\begin{figure*}[h]
\begin{minipage}{18pc}
 \centering
 \centering
 \includegraphics[width=16pc]{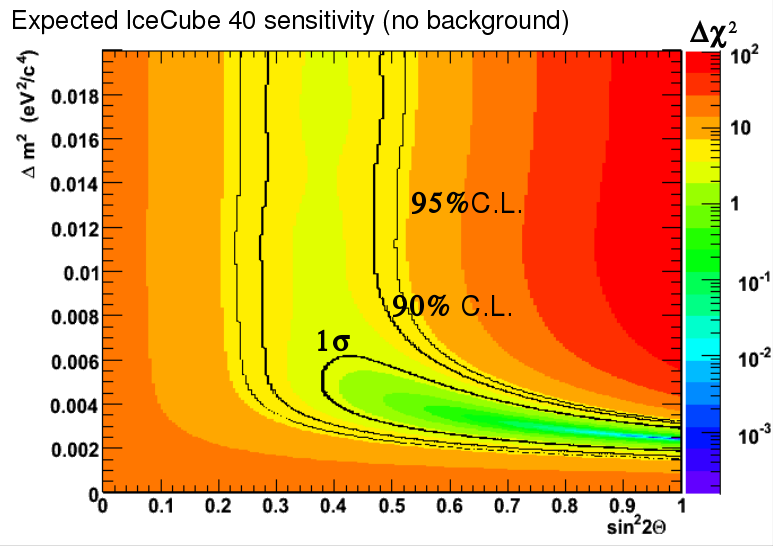}
 \caption{Theoretical sensitivity $\Delta \chi^2 = \chi^2-\chi^2_{\rm min}$ distribution for the IceCube 40-string
   detector based on the selection criteria defined in the presented
   analysis, assuming no remaining background. Contours correspond to
   $\chi^2_{\rm min}$ + 2.3, 4.6, 6.0.} 
   \label{plot:oscillation_sensitivity_IC40}
\end{minipage}\hspace{2pc}%
\begin{minipage}{18pc}
 \centering
 \includegraphics[width=16pc]{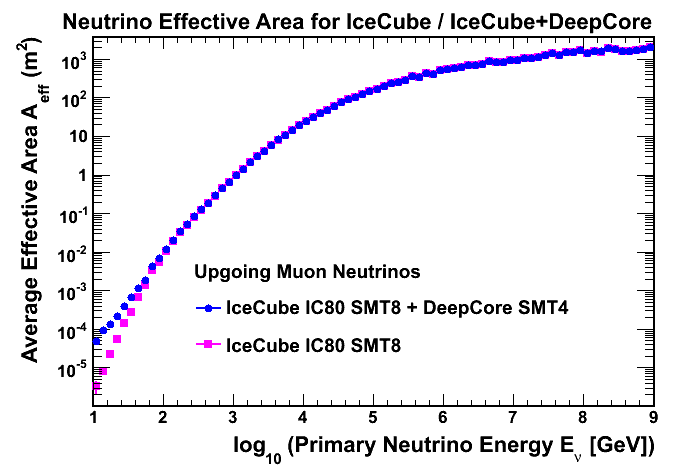}
 \caption{Comparison of preliminary study of effective area ${\rm A}_{\rm eff}$ for the full IceCube
     detector (IC80) and in addition with Deep Core at trigger level. 
     The  addition of Deep Core increases the effective area of the detector
     at low energies significantly.} \label{plot:efficitive_area}

\end{minipage}
\end{figure*}

Based on the selection criteria for this IceCube 22-string analysis we have
evaluated the sensitivity for the IceCube 40-string detector with one year of 
data using a $\chi^2$-test on the track length distribution. 
Figure~\ref{plot:oscillation_sensitivity_IC40} shows the
theoretical sensitivity limits obtained in this way as function of the oscillation 
parameters. The trigger system for the 40-string detector has also been significantly
improved over the 22-string detector through the addition of a string
trigger~\cite{Gross:2007zzb}, which roughly doubles the
vertical muon neutrino candidate events per string.

\section{Conclusions}

We present preliminary results obtained with a subset of IceCube 22-string
dataset collected during 2007 and 2008. The results demonstrate that IceCube is
sensitive to a $\nu_\mu$ energy range in which atmospheric neutrino oscillations become
important. We estimate the sensitivity to oscillation parameters in the
IceCube 40-string dataset and find that we can potentially constrain them, pending the
finalization of the systematic uncertainties associated with the predicted
distributions.
Understanding of this energy region is extremely important also for 
dark matter annihilation signals from the center of the Earth and further
provides the groundwork for Deep Core, which will probe
neutrinos at a similar and even lower energy range. This analysis allows to
test our understanding of these events as part of a physics analysis.
Deep Core, which will significantly improve IceCube's
sensitivity in the energy range below 100~GeV (see also
Figure~\ref{plot:efficitive_area}),
will allow not only to observe muon neutrino disappearance but also
possibly appearance of tau neutrinos as well as to measure oscillation 
effects as a function of the zenith
angle. It has generated further interest in the community with its capability
to potentially resolve the neutrino mass hierarchy~\cite{Mena:2008rh}.

\begin{acknowledgments}
We acknowledge the support from the following agencies: National Science
Foundation-Office of Polar Program, National Science Foundation-Physics
Division, University of Wisconsin Alumni Research Foundation, Department of
Energy, and National Energy Research Scientific Computing Center; Swedish
Research Council, Swedish Polar Research Secretariat, and Knut and Alice
Wallenberg Foundation, Sweden; German Ministry for Education and Research,
Deutsche Forschungsgemeinschaft (DFG), Germany; Fund for Scientific Research
(FNRS-FWO), Flanders Institute to encourage scientific and technological
research in industry (IWT), Belgian Federal Science Policy Office (Belspo);
the Netherlands Organisation for Scientific Research (NWO).
\end{acknowledgments}

\end{document}